\documentclass[11pt,a4paper,pdflatex]{article}
\usepackage{jheppub}
\usepackage{physics}
\usepackage{subcaption}
\usepackage{hyperref}
\usepackage{comment}

\title{Scheme dependence and instability of double-trace deformations for gauge fields in AdS$_5$}

\author[a]{Shuta Ishigaki}
\author[a,b]{Masataka Matsumoto}
\affiliation[a]{%
Department of Physics, Shanghai University, 99 Shangda Road, Shanghai 200444, China
}
\affiliation[b]{Department of Physics, Chuo University, 1-13-27 Kasuga, Bunkyo-ku, Tokyo 112-8551, Japan}

\emailAdd{shutaishigaki@shu.edu.cn}
\emailAdd{mmatsumoto173@g.chuo-u.ac.jp}

\abstract{
In holography, double-trace deformations provide a general framework for deforming boundary field theories.
In particular, they can be utilized to introduce dynamical gauge fields in the boundary theory through double-trace deformations of bulk gauge fields.
In this work, we study this construction in the case where the bulk geometry is asymptotically AdS$_5$, and find that such a system involves tachyon and ghost modes.
This instability originates from the logarithmic behavior of the gauge fields in the vicinity of the AdS boundary, which leads to a scheme-dependent ambiguity in the double-trace deformation.
We investigate this instability by using both analytical and numerical methods in several holographic setups, including bottom-up models and the top-down D3–D7 construction.
}
\begin{document}
\maketitle

\section{Introduction}
The holographic method is one of the powerful tools for analyzing strongly coupled quantum field theory, and condensed matter systems.
In the standard holographic dictionary, a local symmetry in the bulk translates into a global symmetry in the boundary theory.
Therefore, even if one considers a bulk gauge theory, such as a $U(1)$ Maxwell theory, the gauge fields are not dynamically living in the boundary theory.
In the boundary theory, the gauge fields simply act as external sources to the current operators.
However, there is a way to introduce gauge fields to the boundary field theory of the holographic models: double-trace deformations.
In general holographic models, the multi-trace deformations can be considered by adding boundary actions, which changes the boundary conditions of the bulk fields to the mixed boundary conditions, when the bulk mass squared takes a value within a specific range \cite{Klebanov:1999tb,Witten:2001ua}.
Considering a specific double-trace deformation of the current operator, one can impose the Maxwell equation on the boundary gauge fields as a boundary condition.\cite{Marolf:2006nd, Domenech:2010nf}
Note that considering a standard quantization for $p$-form fields, which are dual of the one-form gauge fields, is also an equivalent method for introducing dynamical gauge fields.\cite{Hofman:2017vwr,Grozdanov:2017kyl,DeWolfe:2020uzb,Iqbal:2020lrt}
This method has been utilized to study various phenomena, where the dynamical gauge fields are essential, in the context of AdS/CMT, e.g., including plasmons \cite{Gran:2017jht, Gran:2018iie, Mauri:2018pzq, Baggioli:2019aqf, Baggioli:2019sio}, Friedel oscillations \cite{Faulkner:2012gt}, anyons \cite{Jokela:2013hta,Brattan:2013wya}, Meissner effects in the holographic superconductors \cite{Natsuume:2022kic,Jeong:2023las,Baggioli:2023oxa,Natsuume:2024sic,Natsuume:2024ril}, and hydrodynamics both in equilibrium \cite{Ahn:2022azl,Ahn:2024ozz} and non-equilibrium steady states \cite{Ahn:2025ptl}.

Despite these developments, the method of the double-trace deformation requires special attention when it is applied to systems in an asymptotically AdS$_{5}$ spacetime, i.e, ($3+1$)-dimensional boundary theory.%
\footnote{
    More precisely, this problem will arise in the case of the AdS$_{d+1}$ spacetime with even $d$.
}
Due to a logarithmic divergence in the bare on-shell action of the five-dimensional bulk theory, the results depend on a renormalization scale.
In the standard quantization, this logarithmic divergence can also be related to a renormalization scheme called differential regularization \cite{Freedman:1991tk}.
The similar divergence in the gravitation part is related to the conformal anomaly in $\mathcal{N}=4$ supersymmetric Yang-Mills theory \cite{Henningson:1998gx,Skenderis:2002wp}.
Without the double-trace deformation, this scheme-dependent ambiguity only affects the real part of the spectral functions, so it can be ignored in many cases.
With the double-trace deformation, however, the ambiguity, represented by a dimensionful parameter in the counterterm, influences the mode frequency \cite{Mauri:2018pzq}.
In previous studies, this scale is often chosen as the (inverse) temperature of the system.
However, we find that the system with the double-trace deformation always has an unstable mode with a positive imaginary frequency for any choice of the constant scale in the case of the AdS$_5$ geometry.
As we will discuss in this paper, this instability commonly appears in various setups of the AdS$_5$ holographic models.
Note that this instability originates from the logarithmic dependence of the on-shell action in an asymptotically AdS$_5$ spacetime. Therefore, this issue does not arise in the case of AdS$_4$ spacetime, see Appendix \ref{appendix:SAdS4}.

{
A similar issue of the tachyon was reported in \cite{Faulkner:2012gt}, and it is related to the presence of ghosts and the violation of the unitarity bound, as studied in \cite{Andrade:2011dg,Andrade:2011aa}.%
\footnote{
 The superluminal mode in \cite{Grozdanov:2017kyl} may come from the same origin.
}
}
The relation between the bulk vector mass $m_{A}$ and the scaling dimension $\Delta$ in the AdS$_{d+1}$ spacetime is given by
\begin{equation}
    m_{A}^2 L^2 = (\Delta - 1)(\Delta + 1 -d),
\end{equation}
with the AdS radius  $L$.
For massless gauge fields, it leads to $\Delta_{-} = 1$ and $\Delta_{+} = d - 1$.
The operator associated  with $\Delta_{+}$ is usually interpreted as a current operator, and it always saturates the unitarity bound of spin-1 operators, $\Delta \geq d-1$.\cite{Minwalla:1997ka}
On the other hand, the operator associated with $\Delta_{-}$ may become a photon operator $A_{m}$ under the double-trace deformation.
However, we cannot apply the same unitarity bound since the photon itself is not gauge invariant.
Instead, we need to consider the unitarity bound for $F_{mn} = \partial_{m} A_{n} - \partial_{n} A_{m}$, which is given by $\Delta \geq d-2$.
Since the scaling dimension of $F_{mn}$ is $2$, the unitarity bound is saturated at $d=4$.
If the unitarity bound is violated, it usually leads to a ghost mode which has a negative norm.

{
Briefly speaking, a ghost mode corresponds to a violation of unitarity and typically signals a pathology of the theory, while a tachyonic mode refers to a fluctuation with an effective negative mass squared, leading to exponential growth of perturbations.}
At zero temperature, the absence of the ghost can be checked from the sign of the appropriate inner product.
At finite temperature, on the other hand, it is not so straightforward since the linearized equations are no longer standard Sturm-Liouville problem due to the incoming-wave condition.
Another possible way is to ensure the positivity of the spectral function, corresponding to the positivity of the norm in a Hilbert space, given by the Green's function of an operator $\mathcal{O}$:
\begin{equation}
    0 \leq \rho(\omega) = \Im G^{\rm R}_{\mathcal{O}\mathcal{O}}(\omega),
\end{equation}
which ensures the absence of ghosts.%
\footnote{
    In our convention, we do not need a minus sign in front of the Green's function.
}

Although the relation between the double-trace deformation and ghost modes or the unitarity bound has been discussed in earlier works \cite{Andrade:2011dg,Andrade:2011aa}, the implications of these issues in asymptotically AdS$_5$ spacetimes have not been thoroughly investigated in the context of recent developments in AdS/CMT. 
In this paper, we investigate the scheme-dependent ambiguity associated with the double-trace deformation for gauge fields in asymptotically AdS$_5$ spacetimes and the resulting instability, focusing on physical observables such as response functions and collective excitations.
As explicit examples, we examine the five-dimensional Einstein-Maxwell theory with and without a charge density, and study the quasi-normal modes with respect to the double-trace deformation.
{
We show that a tachyon always exists for any choice of the renormalization scale in these cases.
The analysis of the corresponding residue at $T=0$ indicates that the tachyon is also a ghost.
}%
Additionally, we consider the deformation in the top-down D3-D7 model \cite{Karch:2002sh}, which also has an AdS$_{5}\times S^{5}$ geometry at zero temperature.
In this setup, the system has normal modes as collective excitations.
We explore not only the behavior of these normal modes but also a Sturm-Liouville norm with the double trace deformation, and demonstrate the emergence of a ghost tachyon associated with the scheme-dependent ambiguity.

{
From the perspective of the renormalization group, we can understand the presence of tachyons as a vacuum instability in a theory with a Landau pole.
Since a theory with a Landau pole is UV incomplete and not valid at high energies,
our observation of the ghost tachyon suggests that the theory under the double-trace deformation similarly becomes pathological at high energies.
This reflects the fact that a quantum field theory with $U(1)$ gauge symmetry, such as quantum electrodynamics (QED), is usually UV incomplete, and the double-trace deformation precisely realizes such a $U(1)$ gauge theory in the boundary field theory.
}

This paper is organized as follows.
In section \ref{sec:double-trace}, we review the method of the double-trace deformation in an asymptotically AdS$_5$ spacetime.
We also review that the resulting response function corresponds to the form in the random phase approximation (RPA), incorporating the effect of dynamical photons.
In section \ref{sec:SAdS5}, we analyze the mode frequencies with the double-trace deformation by using the analytic solution in the Schwarzschild AdS$_5$ spacetime.
In section \ref{sec:RNAdS5}, we perform numerical analysis for the charged system described by the Reissner-Nordstr\"om AdS$_5$ black hole.
In section \ref{sec:D3-D7_Minkowski}, we analyze the normal modes in the top-down D3-D7 model with general boundary conditions.
We present conclusion and discussions in section \ref{sec:discussions}.
In appendix \ref{appendix:SAdS4}, we perform the analysis in the Schwarzschild AdS$_{4}$ and show that there is no unstable mode in this case.

\section{The double-trace deformation: a review}\label{sec:double-trace}
In this section, we review the method to introduce the double-trace deformation to make the boundary gauge fields dynamical, based on \cite{Mauri:2018pzq} and \cite{DeWolfe:2020uzb}.
In the following, we focus on the bulk five-dimensional setup.

First of all, we need to consider an appropriate counterterm according to the procedure of holographic renormalization \cite{Skenderis:2002wp}.
Here, we consider a five-dimensional Maxwell theory in the bulk:
\begin{equation}
    S_{\rm bulk} =
    \int \dd[5]{x} \sqrt{-g}\left(
        -\frac{1}{4} F_{\mu\nu} F^{\mu\nu}
    \right),
\end{equation}
where $\mu,\nu$ are indices over ($t, x, y, z, u$).
The background metric is assumed to be an asymptotically AdS$_5$ spacetime in the vicinity of $u=0$, where it behaves as
\begin{equation}
    \dd{s}^2 \approx \frac{- \dd{t}^2 + \dd{\vec{x}}^2 + \dd{u}^2}{u^2},
\end{equation}
where $\vec{x} = (x,y,z)$.
For the gauge fields in the asymptotically AdS$_5$ spacetime, we need to consider
\begin{equation}
	S_{\rm ct} = \int_{\epsilon} \dd[4]{x} \sqrt{-\gamma}\left(
		- \frac{1}{4} \log(\frac{\epsilon}{u_{\rm ct}}) F_{mn} F^{mn}
	\right),
    \label{eq:counterterm}
\end{equation}
where $u=\epsilon$ is a small cutoff, and $u_{\rm ct}$ is an arbitrary scale to fix the (scaling) dimensionality or to represent the ambiguity of the cutoff scale.
$m,n$ are indices for the boundary coordinates $(t,x,y,z)$.
$\gamma_{mn}$ is the induced metric on the slice at $u=\epsilon$.
We obtain the renormalized action for the standard quantization as
\begin{equation}
	S_{\rm std} = \lim_{\epsilon \to 0} S_{\rm reg} + S_{\rm ct},
\end{equation}
where $S_{\rm reg}$ is a regularized on-shell bulk action.
Taking the variation of each term, we obtain
\begin{align}
	\var{S}_{\rm reg} =&
	 - \int_{\epsilon}\dd[4]{x} \sqrt{-g} n_{\mu} F^{\mu\nu} \var{A}_{\mu},\\
	 \var{S}_{\rm ct} =&
	 \int_{\epsilon} \dd[4]{x} \sqrt{-\gamma}
     \log(\frac{\epsilon}{u_{\rm ct}})
     \left(
        \nabla_{m} F^{mn}
    \right)\var{A}_{n},
\end{align}
where $n_{\mu} = - \delta^{u}{}_{\mu}$ is a normal vector to the cutoff hypersurface.
Note that we dropped the surface term that comes from the black-hole horizon according to the standard prescription \cite{Son:2002sd}.
This prescription is justified by considering a correspondence between the holographic Green's functions and the Schwinger-Keldysh formalism.\cite{Herzog:2002pc}
In Fourier space, the asymptotic expansion of massless vector fields in AdS$_{5}$ generally becomes
\begin{equation}
	A_{m} = \alpha_{m}\left(
		1 + \frac{k^2}{2} u^2 \log \frac{u}{u_{\rm ex}}
	\right) + \beta_{m} u^2 + \cdots,
\end{equation}
where $\alpha_{m}$ is a leading coefficient with scaling dimension $1$, and $\beta_{m}$ is a subleading coefficient with scaling dimension $3$.
The constant $u_{\rm ex}$ is also an arbitrary scale to fix the dimensionality, but it does not affect the final result.
In the literature, this scale is often chosen $u_{\rm ex}^2 = 1/\abs{k^2}$.
By substituting this into the renormalized action, we obtain
\begin{equation}
    \var{S}_{\rm std} = \int \frac{\dd[4]{k}}{(2\pi)^4}\left[
        2 \beta^{m}(k)
        + k^2 \left(
            \frac{1}{2}
            + \log\frac{u_{\rm ct}}{u_{\rm ex}}
        \right) \alpha^{m}(k)
    \right]\var{\alpha}_{m}(-k).
    \label{eq:variation_standard}
\end{equation}
We rewrite this as
\begin{equation}
	\var{S}_{\rm std} = \int \frac{\dd[4]{k}}{(2\pi)^4} \Pi^{m}(k) \var{\alpha}_{m}(-k).
\end{equation}
The current-current response function is given by%
\footnote{
{
More precisely, the current density corresponds to $-\Pi^{m}$ in our setup.
}
}
\begin{equation}
	G^{mn} = \fdv{\Pi^{m}}{\alpha_{n}}
    = 2 \fdv{\beta^{m}}{\alpha^{n}}
    + k^2
    \left(
        \frac{1}{2}
        + \log\frac{u_{\rm ct}}{u_{\rm ex}}
    \right) \eta^{mn}
    .
    \label{eq:Green's_function}
\end{equation}
Eq.~(\ref{eq:variation_standard}) implies that the leading term $\alpha_{m}$ plays the role of the external source.
On the other hand, the response is given by the combination of $\alpha_m$ and $\beta_m$, and it depends on the choice of $u_{\rm ct}$.

Next, let us consider the double-trace deformation.
We introduce the following boundary action \cite{Mauri:2018pzq}
\begin{equation}
	S_{\rm fin} = \int \dd[4]{x}
    \left[
		- \frac{1}{4 \lambda} f^{mn} f_{mn} + \alpha_{m} J^{m}_{\rm ext}
	\right],
    \label{eq:Sfin}
\end{equation}
where $\lambda$ is a coupling constant, and $f_{mn} = \partial_{m} \alpha_{n} - \partial_{n} \alpha_{m}$.%
\footnote{
    It is also equivalent to consider the Dirichlet condition on the dual two-form fields \cite{DeWolfe:2020uzb}.
}
Note that the finite part of the counterterm (\ref{eq:counterterm}) can be absorbed into the term proportional to $1/\lambda$ here.
Thus, the coupling $\lambda$ above can be regarded as a renormalized coupling defined at the renormalization scale $M=1/u_{\rm ct}$.
The variation of the boundary action becomes
\begin{equation}
	\var{S}_{\rm fin} =
	\int\dd[4]{x}
	\left[
		\left(
			\frac{1}{\lambda}\nabla_{m} f^{mn} + J^{n}_{\rm ext}
		\right) \var{\alpha}_{n}
		+ \alpha_{m} \var{J}^{m}_{\rm ext}
	\right].
\end{equation}
Together with it, we obtain
\begin{equation}
	\var{S}_{\rm alt}= \var{S}_{\rm std} + \var{S}_{\rm fin}
	=
	\int\dd[4]{x}
    \left[
		\left(
			\frac{1}{\lambda}\nabla_{m} f^{mn} + \Pi^{n} + J^{n}_{\rm ext}
		\right) \var{\alpha}_{m}
		+ \alpha_{m} \var{J}^{m}_{\rm ext}
	\right].
    \label{eq:variation_Salt}
\end{equation}
If the coefficient of $\var{\alpha}_{m}$ vanishes, the variational problem is well-posed in this case as well.
Thus, we determine the source term $J^{m}_{\rm ext}$ from the boundary Maxwell equation
\begin{equation}
	\frac{1}{\lambda} \partial_{m} f^{mn} + \Pi^{n} + J^{n}_{\rm ext} =0.
    \label{eq:boundary_Maxwell_equation}
\end{equation}
Since the variation becomes $\var{S}_{\rm alt} = \int \alpha_{m} \var{J}^{m}_{\rm ext}$, the leading-term coefficient $\alpha_{m}$ now plays the role of the response.

It is also useful to write the expressions in terms of the gauge invariants.
In this paper, we take the momentum of the perturbation fields to be  $k^{m} = (\omega, k_{x}, 0, 0)$.
We take the gauge invariants for the longitudinal and the transverse sectors as
\begin{equation}
	Z_{1} = i\omega A_{x}(u) + i k_{x} A_{t}(u),\quad
	Z_{2} = i\omega A_{y}(u),
    \label{eq:gauge_invariants}
\end{equation}
in Fourier space, respectively.%
\footnote{
    In this paper, we take the convention of the Fourier transformation as
    $\Phi(x) = \int\frac{\dd[d]{x}}{(2\pi)^{d}} e^{i k\cdot x}\Phi(k)$
    or $\Phi(t,x) = \int \frac{\dd{\omega}\dd{k_{x}}}{(2\pi)^2}
    e^{-i \omega t + i k_{x} x} \Phi(\omega, k_{x})$.
}
We omit the perturbation along the $z$-direction since its dynamics is the same as that along the $y$-direction.
We can write each source as
\begin{align}
	J_{\rm ext}^{t} =&
	\frac{-i k_{x}}{k_{x}^2 - \omega^2}
	\left(
		2 Z_{1}^{(S)}
		- (k_{x}^2 - \omega^2)
		Z_{1}^{(L)}\left(
			\frac{1}{\lambda}
			- \log\frac{u_{\rm ct}}{u_{\rm ex}}
			- \frac{1}{2}
		\right)
	\right),\\
	J_{\rm ext}^{x} =&
	\frac{-i \omega}{k_{x}^2 - \omega^2}
	\left(
		2 Z_{1}^{(S)}
		- (k_{x}^2 - \omega^2)
		Z_{1}^{(L)}\left(
			\frac{1}{\lambda}
			- \log\frac{u_{\rm ct}}{u_{\rm ex}}
			- \frac{1}{2}
		\right)
	\right)
	,\\
	J_{\rm ext}^{y} =&
	\frac{1}{-i\omega}\left(
		2 Z_{2}^{(S)}
		- (k_{x}^2 - \omega^2) Z_{2}^{(L)} \left(
			\frac{1}{\lambda}
			- \log\frac{u_{\rm ct}}{u_{\rm ex}}
			- \frac{1}{2}
		\right)
	\right),
\end{align}
where $Z^{(L)}_{\alpha}$ and $Z^{(S)}_{\alpha}$ are the leading and sub-leading coefficients in the asymptotic expansion:
\begin{equation}
	Z_{\alpha} = Z_{\alpha}^{(L)} \left(
		1 + \frac{k_{x}^2 - \omega^2}{2} u^2 \log \frac{u}{u_{\rm ex}}
	\right)
	+ Z_{\alpha}^{(S)} u^2 + \order{u^3},
\end{equation}
respectively.
{
By solving $J^{m}_{\rm ext}=0$, we obtain
\begin{equation}
	\frac{1}{\lambda} =
	 \frac{2}{k_{x}^2 -\omega^2}
	\frac{Z^{(S)}_{\alpha}}{Z^{(L)}_{\alpha}}
	+ \frac{1}{2} + \log\frac{u_{\rm ct}}{u_{\rm ex}}.
    \label{eq:lambda_inverse}
\end{equation}
Note that $Z^{(S)}_{\alpha}/Z^{(L)}_{\alpha}$ depends on the choice of $u_{\rm ex}$.
One can see that varying $u_{\rm ct}$ just shifts $1/\lambda$ by a constant.
This behavior can be understood from the perspective of the renormalization group.\cite{Grozdanov:2017kyl, Hofman:2017vwr,DeWolfe:2020uzb}
The observables depend only on the RG-invariant scale
\begin{equation}
    u_{\star} := u_{\rm ct} e^{-\frac{1}{\lambda}}.
    \label{eq:RG-invariant}
\end{equation}
The coupling in the boundary action can be interpreted as a renormalized coupling fixed by the renormalization point at $M = 1/u_{\rm ct}$: $\lambda = \lambda_{\rm eff}(M)$.
By choosing $u_{\rm ex} = 1/Q$ where $Q = \abs{k}$, we rewrite (\ref{eq:lambda_inverse}) as
\begin{equation}
    \frac{1}{\lambda_{\rm eff}(M)} =
    \left.
    \frac{2}{k_{x}^2 - \omega^2}
	\frac{Z^{(S)}_{\alpha}}{Z^{(L)}_{\alpha}}
	+ \frac{1}{2} 
    + \log\frac{Q}{M}
    \right|_{u_{\rm ex} = \frac{1}{Q}},
\end{equation}
so that we obtain
\begin{equation}
    \frac{1}{\lambda_{\rm eff}(Q)}
    =
    \left.
	\frac{2}{k_{x}^2 - \omega^2}
	\frac{Z^{(S)}_{\alpha}}{Z^{(L)}_{\alpha}}
	+ \frac{1}{2} 
    \right|_{u_{\rm ex} = \frac{1}{Q}}.
\end{equation}
We can also write
\begin{equation}
    \lambda_{\rm eff}(M) =
    \frac{\lambda_{\rm eff}(Q)}{
        1 + \lambda_{\rm eff}(Q) \log(Q/M)
    }.
    \label{eq:running_coupling}
\end{equation}
From this equation, the beta function is obtained \cite{Grozdanov:2017kyl,DeWolfe:2020uzb}
\begin{equation}
    \beta_{\lambda}(M)
    :=
    M \pdv{\lambda_{\rm eff}}{M}
    = \lambda_{\rm eff}(M)^2.
\end{equation}
The results are parallel to those in QED.%
\footnote{
For example, see section 16.3.3 of textbook \cite{Schwartz:2014sze}.
}
The positive beta function implies the existence of the Landau pole, at which the coupling constant diverges.
}



\begin{figure}
    \centering
    \includegraphics[width=0.6\linewidth]{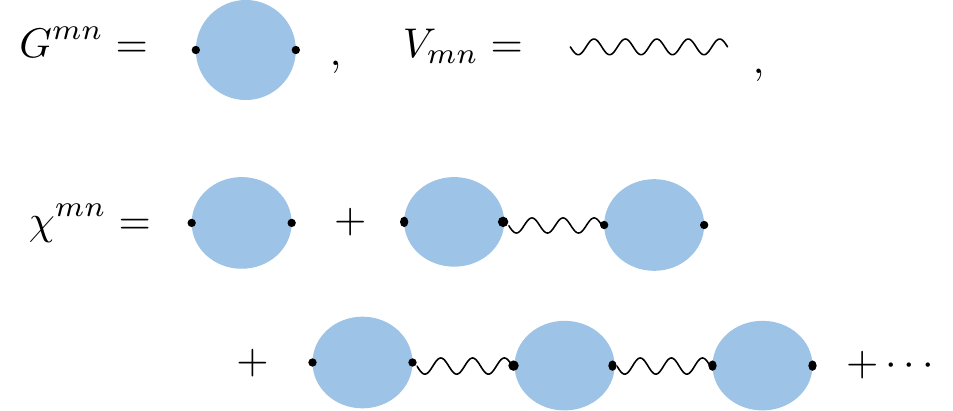}
    \caption{
    Diagrammatic representation of Eq.~(\ref{eq:RPA}).
    The cyan circle for $G^{mn}$ represents the holographic Green's function without dynamical gauge fields.
    {
    The wavy line for $V_{mn}$ denotes the Green's function of free photons.
    }%
    Each vertex brings the coupling constant $\sqrt{\lambda}$.
    }
    \label{fig:RPA_diagram}
\end{figure}

Following to \cite{Mauri:2018pzq}, we can define the (current-current) response function with respect to external gauge fields in systems with the double-trace deformation.
The new response function is related to the original Green's function by a relation similar to that in the random phase approximation (RPA).
Let us now add a gauge-fixing term $\frac{1}{\lambda}\int\dd[4]{x}\frac{1}{2\xi}(\partial_{m} \alpha^{m})^2$ in the R$_\xi$ gauge into eq.~(\ref{eq:Sfin}).
In the Fourier space, we write
\begin{equation}
    S_{\rm fin} = \int\frac{\dd[4]{k}}{(2\pi)^4}
    \left[
        -\frac{1}{2\lambda}\alpha_{m}(k) W^{mn}(k) \alpha_{n}(-k)
        + J^{m}(k)\alpha_{m}(-k)
    \right],
    \label{eq:Sfin_Fourier}
\end{equation}
where
\begin{equation}
    W^{mn}(k) :=
    k^2\left(\eta^{mn} - \frac{\xi - 1}{\xi}\frac{k^{m}k^{n}}{k^2}\right).
\end{equation}
By performing the Gaussian integral for $\alpha^{m}$, we can write
\begin{equation}
    S_{\rm fin}
    =
    - \frac{\lambda}{2}\int\frac{\dd[4]{k}}{(2\pi)^4}
    J^{m}_{\rm ext}(k)V_{mn}(k)
    J^{n}_{\rm ext}(-k),
\end{equation}
where
\begin{equation}
    V_{mn}(k) := \frac{1}{k^2}\left(
        \eta_{mn} - (1-\xi)\frac{k_{m}k_{n}}{k^2}
    \right).
\end{equation}
Note that $V_{mn}$ is the inverse of $W^{mn}$.
Owing to the R$_{\xi}$ gauge, we can invert the matrix $W^{mn}$.
{
The matrix $V_{mn}$ is the (classical) Green's function of free photons in the R$_{\xi}$ gauge.
}
In this gauge, the condition (\ref{eq:boundary_Maxwell_equation}) is written as
\begin{equation}
    - \frac{1}{\lambda} W^{mn} \alpha_{n} + \Pi^{m} + J_{\rm ext}^{m} = 0.
\end{equation}
Solving the equation for $\alpha_{m}$, we can split the boundary values of the gauge fields into two parts:
\begin{equation}
    \alpha_{m} = \lambda V_{mn} \Pi^{n} + \lambda V_{mn} J_{\rm ext}^{n}.
\end{equation}
Comparing this result with (3.9) in \cite{Mauri:2018pzq}, we rewrite it as
\begin{equation}
    J_{\rm ext}^{m} = \frac{1}{\lambda} W^{mn} \alpha_{n}^{\rm ext},
    \label{eq:rel_Jext_Aext}
\end{equation}
then obtain
\begin{equation}
    \alpha_{m} = \alpha_{m}^{\rm ext} + \lambda V_{mn} \Pi^{n},
    \label{eq:relation_with_external_gauge_fields}
\end{equation}
where $\alpha_{n}^{\rm ext}$ is interpreted as external $U(1)$ gauge fields.%
\footnote{
    In \cite{Mauri:2018pzq}, they regarded $J_{\rm ext}^{m}$ as $\Pi^{m}$ in some places but we believe $J_{\rm ext}^{m}$ should be related to $\alpha_{m}^{\rm ext}$.
    We can obtain the same results even with our interpretation.
}
Substituting (\ref{eq:rel_Jext_Aext}) into (\ref{eq:Sfin_Fourier}), we obtain the Maxwell action with $\alpha_{m}$ replaced by $\alpha_{m} + \alpha_{m}^{\rm ext}$ apart from a quadratic term in $\alpha_{m}^{\rm ext}$.
This is a natural consequence of the interpretation of $\alpha_{n}^{\rm ext}$.
Now, we define the response function $\chi^{mn}$ as
\begin{equation}
    \chi^{mn}:= \fdv{\Pi^{m}}{\alpha_{n}^{\rm ext}}.
\end{equation}
Note that the one-point function with respect to $\alpha_{m}^{\rm ext}$ is given by
\begin{equation}
    \fdv{S_{\rm alt}}{\alpha_{m}^{\rm ext}}
    = \alpha_{n} \fdv{J_{\rm ext}^{n}}{\alpha_{m}^{\rm ext}}
    = - \frac{1}{\lambda} W^{mn} \alpha_{n}^{\rm ext} - \Pi^{m},
\end{equation}
where the first term is ignorable since it originates from the quadratic term of $\alpha_{n}^{\rm ext}$ in the action.
Thus $\Pi^{m}$ is still the main contribution to the one-point function, and $\chi^{mn}$ is the corresponding two-point function.
Using the chain rule together with (\ref{eq:relation_with_external_gauge_fields}), we obtain the RPA-like formula
\begin{equation}
    \chi =
    \left(
        I - \lambda G V
    \right)^{-1} G,
    \label{eq:RPA}
\end{equation}
where $G$ is defined by (\ref{eq:Green's_function}).
The diagrammatic representation is shown in figure \ref{fig:RPA_diagram}.
From the Ward identity, the Green's function must be transverse: $k_{m}G^{mn}=0$.
Using the projection matrix, we assume that the Green's function has the form
\begin{equation}
    G^{mn}(k) = (k^2 \eta^{mn} - k^{m}k^{n})\Pi(k),
\end{equation}
where $\Pi(k)$ is a vacuum polarization function.
Substituting this form of the Green's function, we obtain the $\xi$-independent result
\begin{equation}
    \chi^{mn}(k) = (k^2 \eta^{mn} - k^{m}k^{n}) \frac{\Pi(k)}{1 - \lambda \Pi(k)}.
\end{equation}
The mode frequencies obtained by solving $J_{\rm ext}^{m}=0$ or $\alpha_{m}^{\rm ext}=0$ correspond to the poles of this response function.
For $k_{m} = (\omega, 0, 0,0)$, we obtain
\begin{equation}
    \chi^{xx}(\omega) = \frac{-\omega^2 \Pi(\omega)}{1 - \lambda \Pi(\omega)}
    = \frac{G^{xx}(\omega)}{1 + \frac{\lambda}{\omega^2} G^{xx}(\omega)}.
    \label{eq:response_function_chi}
\end{equation}
We can write its imaginary part as
\begin{equation}
    \Im\chi^{xx}(\omega) = \frac{\Im G^{xx}(\omega)}{\abs{1+\frac{\lambda}{\omega^2}G^{xx}(\omega)}^2},
    \label{eq:rel_spectral_func}
\end{equation}
so the sign of the spectral function is unchanged from the original one.
For later use, we also define the AC conductivity, taking into account dynamical electromagnetic fields, as
\begin{equation}
    \sigma_{\rm em}(\omega) := \frac{\chi^{xx}(\omega)}{i\omega}.
    \label{eq:sigma_em}
\end{equation}

\section{Examples in the Schwarzschild \texorpdfstring{AdS$_5$}{AdS5} spacetime}\label{sec:SAdS5}
In this section, we show the presence of unstable modes in the Schwarzschild AdS$_5$ spacetime with the double-trace deformation.
In this case, we can use an analytic solution for the vector perturbation fields.

We employ the five-dimensional Einstein-Maxwell theory with a negative cosmological constant in the bulk:
\begin{equation}
	S = \int\dd[5]{x}\sqrt{-g}\left[
		R - \frac{6}{L^2} - \frac{1}{4} F_{\mu\nu} F^{\mu\nu}
	\right].
    \label{eq:Einstein-Maxwell_action}
\end{equation}
Here, we consider the case of vanishing charge density, so there is no backreaction.
The metric solution is given by the Schwarzschild AdS$_5$:
\begin{equation}
	\dd{s}^2 = \frac{-f(u) \dd{t}^2 + \dd{\vec{x}}^2}{u^2} + \frac{\dd{u}^2}{f(u) u^2},\quad
	f(u) = 1 - \frac{u^4}{u_{0}^4}.
\end{equation}
We set $L=1$ for simplicity.
The black hole horizon is at $u=u_{0}$ and the Hawking temperature is given by $T = 1/(u_{0}\pi)$.
Let us consider small perturbations
\begin{equation}
	A = a_{t}(t,x,u) \dd{t} + a_{x}(t,x,u) \dd{x} + a_{y}(t,x,u) \dd{y}. 
\end{equation}
The equations of motion can be written in terms of the gauge invariants (\ref{eq:gauge_invariants}) as
\begin{subequations}
\label{eq:perturbation_equations_SAdS5}
\begin{align}
	0=&
	Z_{1}''(u) + \left(
		- \frac{1}{u}
		+ \frac{\omega^2 f'(u)}{(\omega^2 - k_{x}^2 f(u))f(u)}
	\right) Z_{1}'(u) + \frac{\omega^2 - k_{x}^2 f(u)}{f(u)^2} Z_{1}(u),\\
	0 =&
	Z_{2}''(u) + \left(
		- \frac{1}{u} + \frac{f'(u)}{f(u)}
	\right) Z_{2}'(u) + \frac{\omega^2 - k^2 f(u)}{f(u)^2} Z_{2}(u),
\end{align}
\end{subequations}
in the Fourier space.

For $k_{x}=0$, the two equations coincide, and the solution is written in terms of the hypergeometric function \cite{Nunez:2003eq, Kovtun:2005ev, Myers:2007we}.
In the following, we omit the subscript $\alpha=1,2$.
The incoming-wave solution is given by
\begin{equation}
\begin{aligned}
	Z(u) =&
	\mathcal{C}\times
    \left(
		1-\frac{u_{0}^2}{u^2}
	\right)^{-i \frac{\tilde{\omega}}{4}}
	\left(
		1 + \frac{u_{0}^2}{u^2}
	\right)^{-\frac{\tilde{\omega}}{4}}\\
	&\times {}_2F_1\left(
		1 - 
		\frac{(1+i) \tilde{\omega}}{4}
		,
		-\frac{(1+i) \tilde{\omega}}{4}
		;
		1 - \frac{i \tilde{\omega}}{2}
		;
		\frac{1}{2} \left(1-\frac{u_{0}^2}{u^2}\right)
	\right),
\end{aligned}
\end{equation}
where $\mathcal{C}$ is a normalization constant, ${}_2F_1$ is the hypergeometric function, and $\tilde{\omega} = u_{0} \omega$.
From the expansion in the vicinity of $u=0$, we obtain
\begin{align}
	\frac{Z^{(S)}}{Z^{(L)}} =
	\frac{i \omega}{{4 u_{0}}} \left[
		2 + i \tilde{\omega}  \left(
			H_{\frac{1-i}{4} \tilde{\omega} }
			+ H_{- \frac{1+i}{4} \tilde{\omega} }
		\right)
	\right]
	+ \frac{\omega^2}{4} \left(
			1-\log\frac{2 u_{\rm ex}^2}{u_{0}^2}
		\right),
\end{align}
where $H_{z}$ denotes the harmonic number.
One can also write it in terms of the digamma function $\psi(z)$ by using $H_{z} = \psi(z+1)+\gamma_{\rm E}$.
The retarded Green's function in the standard quantization is given by
\begin{equation}
\begin{aligned}
    G^{xx} =& \frac{\Pi^{x}}{\alpha_{x}}
    =
    2 \frac{Z^{(S)}}{Z^{(L)}} - \omega^2\left(
        \frac{1}{2} + \log\frac{u_{\rm ct}}{u_{\rm ex}}
    \right)\\
    =&
    \frac{i \omega}{{2 u_{0}}} \left[
		2 + i \tilde{\omega}  \left(
			H_{\frac{1-i}{4} \tilde{\omega} }
			+ H_{- \frac{1+i}{4} \tilde{\omega} }
		\right)
	\right]
    + \frac{\omega^2}{2} \log\frac{u_{0}^2}{2 u_{\rm ct}^2},
\end{aligned}
\end{equation}
where the poles are at $\omega = 2 n (\pm1-i)\pi T$ for $n = 1,2,3,\cdots$ \cite{Nunez:2003eq,Myers:2007we}.

\begin{figure}[htbp]
	\centering
	\includegraphics[width=0.66\linewidth]{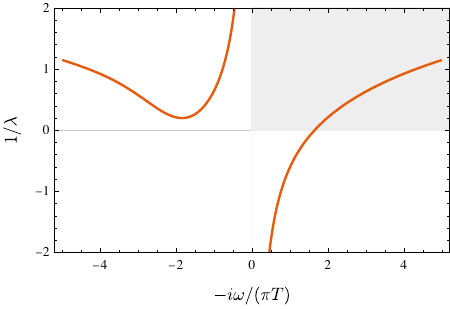}
	\caption{
		$1/\lambda$ as a function of $-i\omega$ with $u_{\rm ct} = 2^{-1/2}u_{0}$ in the SAdS$_{5}$ spacetime.
        The gray region indicates the first quadrant: $-i\omega\geq 0$ and $1/\lambda \geq 0$.
		The curve in this region implies the presence of the unstable modes for the corresponding choice of $\lambda$.
        Note that we set $k_{x} =0$.
	}
	\label{fig:lambda_SAdS5}
\end{figure}

\begin{figure}[htbp]
	\centering
	\includegraphics[width=0.5\linewidth]{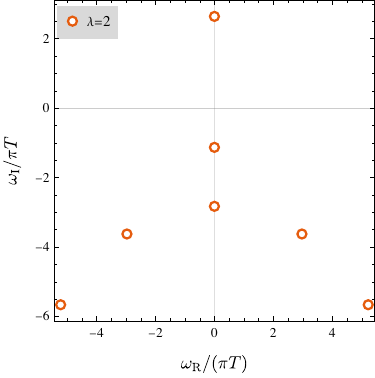}
	\caption{
        Locations of the quasinormal modes for $\lambda=2$ with $u_{\rm ct} = 2^{-1/2}u_{0}$.
        Note that $\omega=\omega_{\rm R} + i \omega_{\rm I}$.
	}
	\label{fig:QNMs_SAdS5}
\end{figure}

With the double-trace deformation, the new source term is given by
\begin{equation}
\begin{aligned}
	i\omega \frac{J_{\rm ext}}{Z^{(L)}}
	=&
	-
	\frac{i \omega}{{2u_{0}}} \left[
		2 + i \tilde{\omega}  \left(
			H_{\frac{1-i}{4} \tilde{\omega} }
			+ H_{- \frac{1+i}{4} \tilde{\omega} }
		\right)
	\right]
	- \omega^2 \left(
		\frac{1}{\lambda} + \frac{1}{2}\log \frac{u_{0}^2}{2 u_{\rm ct}^2}
	\right).
\end{aligned}
\end{equation}
The mode equation is given by $\tilde{J}_{\rm ext} := J_{\rm ext}/Z^{(L)} =0$.
As mentioned in \cite{Mauri:2018pzq}, the choice of $u_{\rm ct}$ now affects the mode frequency.
Solving the equation for $1/\lambda$, we obtain
\begin{equation}
	\frac{1}{\lambda} =
	\frac{1}{2 i \tilde{\omega}} \left[
		2 + i \tilde{\omega}  \left(
			H_{\frac{1-i}{4} \tilde{\omega} }
			+ H_{- \frac{1+i}{4} \tilde{\omega} }
		\right)
	\right]
	+ \frac{1}{2} \log\frac{2 u_{\rm ct}^2}{u_{0}^2}.
\end{equation}
Note that $\lambda$ is real by definition, so this equation is meaningful only when its right-hand side is real.
We show $1/\lambda$ as a function of $-i\omega = \omega_{\rm I}$ in figure \ref{fig:lambda_SAdS5}.
Here, we set $u_{\rm ct} = 2^{-1/2} u_{0}$.
Remarkably, $1/\lambda$ takes both positive and negative values in the region $-i\omega > 0$.
This behavior implies that the unstable mode always exists for any choice of $\lambda$.
Figure \ref{fig:QNMs_SAdS5} shows the locations of the quasinormal modes for $\lambda=2$.
With the current choice of $u_{\rm ct}$, there is actually an unstable mode with $\omega/(\pi T) \simeq + 2.6 i$.

For large $\omega_{\rm I}$, $1/\lambda$ grows logarithmically.
Thus, there is an unstable mode for any small $\lambda$.
The large-$\abs{\omega}/(\pi T)$ behavior is actually equivalent to the result at $T=0$, i.e., in the AdS$_{5}$ spacetime.
At $T=0$, the appropriate solution is given by
\begin{align}
	Z_{\alpha} =
	\mathcal{C} \times
	\begin{cases}
		u H^{(1)}_{1}(u \sqrt{\omega^2 - k_{x}^2}) & \omega^2 - k_{x}^2 >0,~\omega \geq 0,\\
		u H^{(2)}_{1}(u \sqrt{\omega^2 - k_{x}^2}) & \omega^2 - k_{x}^2 >0,~\omega <0,\\
		u K_{1}(u \sqrt{k_{x}^2 - \omega^2}) & \omega^2 - k_{x}^2 \leq 0,
	\end{cases}
\end{align}
where $H^{(1)}_{n}$ and $H^{(2)}_{n}$ are the Hankel functions of the first and second kind, respectively.
$K_{n}$ is the modified Bessel functions of the second kind.
Without the double-trace deformation, it leads to the CFT-type Green's function for the spacelike momentum:
{
\begin{equation}
	G^{xx}(k) =
        k^2 \left(
		\gamma_{\rm E} +\log u_{\rm ct} + \frac{1}{2}\log \frac{\abs{k}}{2}
	\right).
\end{equation}
}
The $\log\abs{k}$ term corresponds to $1/\abs{x}$ in position space.
With the double-trace deformation, we obtain the external source for the spacelike momentum as
{
\begin{equation}
	\tilde{J}_{\rm ext} =
	\frac{k^2}{\lambda}
	- k^2\left(
		\gamma_{\rm E} + \log u_{\rm ct} + \log\frac{\abs{k}}{2}
	\right).
\end{equation}
}
A similar result was found in \cite{Hofman:2017vwr}.
By solving $\tilde{J}_{\rm ext}=0$, we obtain
{
\begin{equation}
	\frac{1}{\lambda} =
	\gamma_{\rm E} + \log u_{\rm ct} + \log\frac{\abs{k}}{2}.
    \label{eq:lambda_inverse_AdS5}
\end{equation}
}
In this case, the logarithmic dependence on $\abs{k}$ is clear.
{
From this result, the mode frequency is obtained
\begin{equation}
    \abs{k_{\star}} := \abs{k} =
    2 e^{1/\lambda - \gamma_{\rm E}}/u_{\rm ct}
    = 2 e^{-\gamma_{\rm E}}/u_{\star},
\end{equation}
where $u_{\star}$ is the RG-invariant scale given by (\ref{eq:RG-invariant}).
Since $k$ is spacelike, this is a tachyon.
The mode frequency, in which $\lambda$ diverges, satisfies
\begin{equation}
    \abs{k} = 2 e^{-\gamma_{\rm E}}/u_{\rm ct}, \label{eq:diverge_momentum}
\end{equation}
at zero temperature.
By using (\ref{eq:response_function_chi}), we obtain the new response function as
\begin{equation}
    \chi^{xx}(k) =
    \frac{
        k^2(\gamma_{\rm E} + \log(\abs{k}u_{\rm ct}/2))
    }{
        1 - \lambda (\gamma_{\rm E} + \log(\abs{k}u_{\rm ct}/2))
    },
\end{equation}
whose expansion around $k^2 = k_{\star}^2$ is given by
\begin{equation}
    \chi^{xx}(k) =
    \frac{
        - 2 k_{\star}^4/\lambda^2
    }{
        k^2 - k_{\star}^2
    } + \order{(k^2 - k_{\star}^2)^{0}}.
\end{equation}
The residue of the above expansion implies that this mode is a ghost.
If we replace $k^2$ with $-\omega^2$, the result is consistent with the result in Section \ref{sec:D3-D7_Minkowski}.
}

From eq.~\eqref{eq:lambda_inverse_AdS5}, one can easily estimate the frequency of the unstable mode. For small $\lambda$, the condition \eqref{eq:lambda_inverse_AdS5} becomes $1/\lambda \sim \frac{1}{2}\log u_{\rm ct}^2 k^2$, so that $u_{\rm ct}^2 k^2 \sim e^{2/\lambda}$.
This implies that we find the unstable mode at the frequency $-i \omega \gg 1/u_{\rm ct}$ for smaller $\lambda$. Even for larger $\lambda$, the frequency becomes $- i u_{\rm ct}\omega \sim 2 e^{-\gamma_{\rm E}} >1$, independent of $\lambda$ as in \eqref{eq:diverge_momentum}. Therefore, the unstable mode is expected to appear at a purely imaginary frequency larger than the renormalization scale; $-i \omega > 1/u_{\rm ct} = M$. This is also the case for finite temperature. For example, with the parameter choice in figure \ref{fig:QNMs_SAdS5}, we obtain
{
\begin{equation}
     \frac{\omega_{\rm I}}{\pi T} \sim
     2 e^{1/\lambda - \gamma_{\rm E}}
     \frac{u_{0}}{u_{\rm ct}}
     = 2 \sqrt{2 e} e^{-\gamma_{\rm E}}
     \approx 2.6
     \label{eq:predict}
\end{equation}
}%
which roughly agrees with the numerical result.


\section{Backreacted finite-density setup}\label{sec:RNAdS5}
In this section, we show that the unstable mode still exists in the case of the backreacted setup with finite density.

The model (\ref{eq:Einstein-Maxwell_action}) admits the Reissner-Nordstr\"om AdS$_{5}$ spacetime solution given by
\begin{gather}
	\dd{s}^2 = \frac{-f(u)\dd{t}^2 + \dd{x}^2 + \dd{y}^2 + \dd{z}^2}{u^2} + \frac{\dd{u}^2}{u^2 f(u)},\\
	f(u) = 1 - \frac{u^4}{u_{0}^4}
	- \frac{\mu^2 u_{0}^2 }{3}\left(
		1 - \frac{u^6}{u_{0}^6}
	\right),\quad
	A = \mu \left(
		1 - \frac{u^2}{u_{0}^2}
	\right) \dd{t},
\end{gather}
where $\mu$ is a chemical potential.
The charge density is given by $\rho = u_{0}^2 \mu$.
The Hawking temperature is given by
\begin{equation}
    T = \frac{1 - \mu^2 u_{0}^2/6}{u_{0} \pi}.
\end{equation}
We now consider linear perturbations around this background solution.
The ansatz for the longitudinal sector is given by
\begin{equation}
\begin{aligned}
	g_{\mu\nu}\dd{x}^{\mu}\dd{x}^{\nu}
	\to&
	\bar{g}_{\mu\nu}\dd{x}^{\mu}\dd{x}^{\nu}\\
	&+ \epsilon\left(
		h_{tt} \dd{t}^2 + 2 h_{tx} \dd{t}\dd{x} + h_{xx} \dd{x}^{2}
		+ h_{yy} \dd{y}^2 + h_{zz} \dd{z}^2
	\right),\\
	A \to&
	\bar{A} + \epsilon a_{t} \dd{t} + \epsilon a_{x} \dd{x},
\end{aligned}
\end{equation}
and the transverse sector is
\begin{equation}
\begin{aligned}
    g_{\mu\nu}\dd{x}^{\mu}\dd{x}^{\nu}
	\to&
	\bar{g}_{\mu\nu}\dd{x}^{\mu}\dd{x}^{\nu}
	+ 2 \epsilon h_{ty} \dd{t} \dd{y}
    + 2 \epsilon h_{xy} \dd{x}\dd{y},\\
    A \to&
	\bar{A} + \epsilon a_{y} \dd{y},
\end{aligned}
\end{equation}
where $\epsilon$ is a small bookkeeping parameter.
Since these sectors are decoupled, we can analyze them separately.
$\bar{g}$ and $\bar{A}$ denote the background solutions but we will omit this notation when there is no potential confusion.
We assume that all the perturbation fields depend only on $t, x$ and $u$ coordinates.
As in the SAdS$_{5}$ case, we can write the linearized equations of motion in terms of the gauge invariants.
Note that we now need to take into account the diffeomorphism invariance.\cite{Kovtun:2005ev}
The gauge invariants we take are
\begin{align}
	Z_{0} =& \frac{g^{xx}}{\omega^2}\left(
		\omega^2 h_{xx} + 2 \omega k_{x}  h_{tx} + k_{x}^2 h_{tt}
		- \left(
			\omega^2 + k_{x}^2 \frac{\partial_{u} g_{tt}}{\partial_{u} g_{yy}}
		\right) h_{yy}
	\right),\\
	Z_{1} =& i \omega a_{x} + i k_{x} a_{t} - i k_{x} \frac{\partial_{u} A_{t}}{\partial_{u} g_{yy}} h_{yy},\\
    Z_{2} =& i \omega a_{y},\\
    Z_{3} =& i \omega h_{xy} + i k_{x} h_{ty}.
\end{align}
For $k_{x}=0$, the equations are decoupled.
The equation of $Z_{\alpha}$ for $\alpha=1,2$ associated with the gauge fields is given by
\begin{equation}
	0 =
	Z_{\alpha}''(u)
	+
	\frac{\left(u^4 A_{t}'(u)^2+18 f(u)-24\right)}{6 u f(u)} Z_{\alpha}'(u)
	+
	\frac{\left(\omega ^2-u^2 f(u) A_{t}'(u)^2\right)}{f(u)^2} Z_{\alpha}(u).
\end{equation}
In this case, we need to solve the above equations numerically.
To obtain physical results, we impose an incoming-wave condition on the gauge invariants at the black hole horizon.

\begin{figure}[htbp]
	\centering
	\includegraphics[width=0.6\linewidth]{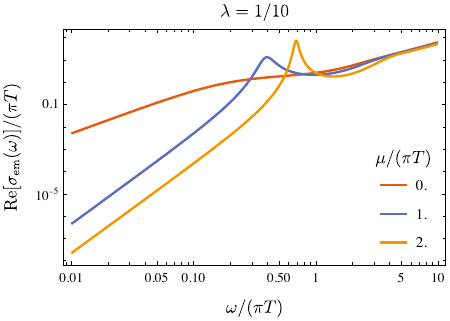}
	\caption{
		Real part of the AC conductivity with dynamical electromagnetic fields for $\mu/(\pi T) = 0, 1.0$ and $2.0$.
        We set $\lambda = 1/10$ and $u_{\rm ct}= 2^{-1/2} u_{0}$.
        Note that the plot is shown in the log-log scale.
	}
	\label{fig:conductivity_em}
\end{figure}

To verify the validity of our analysis, we show the AC conductivity with dynamical electromagnetic fields, which is defined by (\ref{eq:sigma_em}), as a function of the frequency in figure \ref{fig:conductivity_em}.
Here, we choose $u_{\rm ct} = 2^{-1/2} u_{0}$, and $\lambda = 1/10$.
The behavior of our results is consistent with those obtained in \cite{Mauri:2018pzq}.
One can see the plasmon-like peaks at finite $\omega$ for non-zero $\mu$.
As expected from (\ref{eq:rel_spectral_func}), the real part of the conductivity takes positive values for positive $\omega$.

\begin{figure}[htbp]
	\centering
	\includegraphics[width=0.6\linewidth]{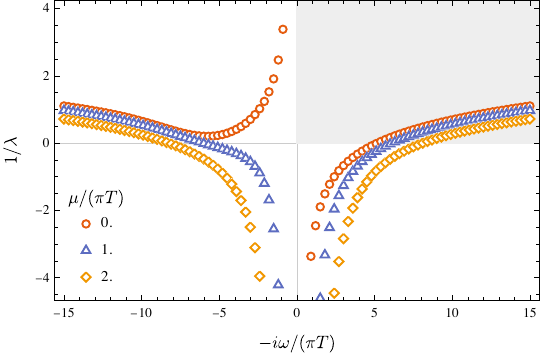}
	\caption{
		$1/\lambda$ as a function of $-i\omega$ with $u_{\rm ct}=2^{-1/2}u_{0}$ for $\mu/(\pi T) = 0, 1, 2$ in the RN-AdS$_{5}$ spacetime.
        Here, we set $k_{x}=0$.
	}
	\label{fig:lambda_RNAdS5}
\end{figure}

Figure \ref{fig:lambda_RNAdS5} shows $1/\lambda$ as a function of $-i\omega$ for various $\mu$.
At $\mu=0$, the system is reduced to one in the SAdS$_{5}$ spacetime, so the result is the same as that shown in figure \ref{fig:lambda_SAdS5}.
Again, we observe that $1/\lambda$ takes positive values in $-i\omega >0$, and it is not bounded from above as in the case of the SAdS$_5$ spacetime.
As $\mu$ increases, $1/\lambda$ tends to decrease.
The large $\omega_{\rm I}$ behavior of $1/\lambda$ is still $\log \abs{\omega_{\rm I}}$.
These results indicate that the system with backreaction is also unstable for any choice of the constant scale $u_{\rm ct}$.

\section{Normal modes in the Minkowski embedding}\label{sec:D3-D7_Minkowski}
In this section, we analyze the worldvolume gauge fields in the D3-D7 model \cite{Karch:2002sh}.
We focus on the zero-temperature case here.
At zero temperature, the D3-branes generate an AdS$_{5}\times S^{5}$ spacetime after taking the throat limit.
The separation between the D3-branes and the D7-brane is interpreted as a quark mass $m_{q}$.
For vanishing charge density, the probe D7-brane can take an embedding that does not fall into the Poincar\'e horizon.
This type of the embeddings is called a \textit{Minkowski embedding}.
In this case, the fluctuations of the D7-worldvolume fields have normal modes.\cite{Kruczenski:2003be}
With a vanishing Dirichlet condition at the boundary, these modes are understood as the excitations of ``mesons.''
See the review of \cite{Erdmenger:2007cm} for further details.
On the other hand, it is mathematically possible to impose Neumann or mixed conditions at the boundary.
We check the validity of such a boundary condition and study the corresponding normal modes from the viewpoint of the norm and the presence of tachyons.

The bulk action of the probe brane is given by the following Dirac-Born-Infeld (DBI) action:
\begin{equation}
    S_{\rm D7} = -\tau_{7}\int\dd[8]{\xi}\sqrt{-\det(g_{ab} + (2\pi\alpha')F_{ab})} +S_{\rm WZ},
    \label{eq:DBI_action}
\end{equation}
where $\tau_{7}$ is the tension of the D7-brane, $\xi^{a}$ are worldvolume coordinates, and $F_{ab}=\partial_{a} A_{b}-\partial_{b} A_{a}$ is the field strength of the worldvolume $U(1)$ gauge fields.
$g_{ab}$ is an induced metric given by
\begin{equation}
    g_{ab} = G_{MN} \partial_{a} X^{M} \partial_{b} X^{N},
\end{equation}
where $X^{M}$ are target space coordinates including the embedding functions, and $G_{MN}$ is the metric of the ten-dimensional AdS$_{5} \times S^{5}$ background given by
\begin{equation}
    \dd{s}_{10}^2 =
    \frac{-\eta_{ij} \dd{x}^{i} \dd{x}^{j} + \dd{u}^2}{u^2}
    + \dd{\Omega}_{5}^2,
\end{equation}
where $\dd{\Omega}_{n}^2$ is the metric of the $n$-sphere.
The five-sphere metric can be decomposed as
\begin{equation}
    \dd{\Omega}_{5}^2 = \dd{\theta}^2 + \sin^2\theta \dd{\varphi}^2 + \dd{\Omega_{3}}^2,
\end{equation}
where ($\theta, \varphi$) are angular coordinates.
$S_{\rm WZ}$ in (\ref{eq:DBI_action}) denotes the Wess-Zumino term but we can neglect it in our setup.
The appropriate counterterms are given by \cite{Karch:2005ms}.
For our purpose, considering (\ref{eq:counterterm}) is sufficient.

We fix the worldvolume coordinate as $\xi^{a} = (x^{i},u,\Omega_{3})$, so $\theta$ and $\varphi$ represent the embedding of the brane.
We focus on the vacuum state without background gauge fields: $A=0$.
In this case, we can set $\varphi = 0$ without loss of generality by virtue of rotational symmetry in the $\varphi$ direction.
By solving the equation of motion, the vacuum solution of the embedding function is given by
\begin{equation}
    \theta(u) = \arcsin( m_{q} u ).
\end{equation}

We now consider the small perturbations
\begin{equation}
    \sin\theta \to \sin\theta(u) + \varepsilon m_{q} u \Phi(x^i,u),\quad
    \varphi \to \varepsilon \varphi(x^i,u),\quad
    A \to \varepsilon A_{i}(x^i,u) \dd{x}^{i},
\end{equation}
where $\varepsilon$ is a small bookkeeping parameter.
Up to quadratic terms of the perturbations, the five-dimensional effective action can be written as
\begin{equation}
    S^{(2)}_{\rm D7} = \mathcal{N}
    \int\dd[5]{x}\sqrt{-\det g_{\mu\nu}}
    \cos^3\theta(u)
    \left(
        -\frac{1}{4} F^2
        - \frac{m_{q}^2 u^2}{2}(\partial \Phi)^2
        - \frac{m_{q}^2 u^2}{2}(\partial \varphi)^2
    \right),
    \label{eq:Seff2}
\end{equation}
where $g_{\mu\nu}$ is a five-dimensional part of the induced metric, and $\mathcal{N} = (2\pi^2)\tau_{7}$.
Hereinafter, we set $\mathcal{N}=1$ for simplicity.
The integration interval of $u$ is $0<u\leq 1/m_{q}$.
Note that we dropped total derivative terms to obtain the above effective action.

The equations of motion for the vector and (pseudo)scalar perturbations are the same in this case.
In the following, we focus on the transverse sector of the vector fields.
In the Fourier space, the linearized equation of motion becomes
\begin{equation}
	0 = \frac{-k^2}{1 -m_{q}^2 u^2} Z(u)
	- \frac{1 + 3 m_{q}^2 u^2}{u (1 - m_{q}^2 u^2)} Z'(u)
	+ Z''(u).
    \label{eq:linearized_eom_Minkowski}
\end{equation}
We can regard this equation as a Sturm-Liouville (SL) problem, whose equation has the form
\begin{equation}
	\dv{u}(p(u)\dv{u} Z(u)) +(\lambda w(u) - q(u))Z(u) = 0,
\end{equation}
where
\begin{equation}
	\lambda = - k^2,\quad
	p(u) = \frac{(1 - m_{q}^2 u^2)^2}{u},\quad
    q(u) = 0,\quad
	w(u) = \frac{1-m_{q}^2 u^2}{u}.
    \label{eq:SL_coeff}
\end{equation}
In the following, we set $k_{x}=0$ for simplicity but we can always recover $k_{x}$ since the system preserves Lorentz invariance at $T=0$.
The solution of eq.~(\ref{eq:linearized_eom_Minkowski}) that is regular at $u=1/m_{q}$  is given by \cite{Kruczenski:2003be, Ishigaki:2023dss}
\begin{equation}
	Z(u) = \mathcal{C}\times
	{}_{2}F_{1}\left(
		\alpha_{+}
		,~
        \alpha_{-}
		, ~
		2 ;~
		1-m_{q}^2 u^2
		\right),
\end{equation}
where
\begin{equation}
    \alpha_{\pm} := \frac{1}{2} \pm \frac{1}{2}
    \sqrt{1+\frac{\omega^2}{m_{q}^2}},\quad
    \alpha_{0} :=
    \frac{\alpha_{+} - \alpha_{-}}{2}
    =
    \frac{1}{2}\sqrt{1+\frac{\omega^2}{m_{q}^2}}.
\end{equation}
Meanwhile, another solution that is singular at $u=1/m_{q}$ is given by
\begin{equation}
\begin{aligned}
	\tilde{Z}(u) =&
	\;\tilde{\mathcal{C}} m_{q}^2 u^2 \times {}_{2} F_{1}\left(
		1 + \alpha_{-}, 1 + \alpha_{+}, 2, m_{q}^2 u^2
	\right).
\end{aligned}
\end{equation}
This solution satisfies the vanishing Dirichlet condition at $u=0$ but does not automatically satisfy the regularity condition at $u=1/m_{q}$.
The IR regularity condition for the singular solution is equivalent to the UV vanishing condition for the regular solution.
In the vicinity of $u=0$, the asymptotic expansion of the solutions can be written as
\begin{equation}
	Z = Z^{(0)} \left(
		1 - \frac{\omega^2 u^2}{2} \log(m_{q} u)
	\right)
	+
	Z^{(\nu)} u^2 + \cdots.
\end{equation}
For the regular solution, we obtain
\begin{align}
	Z^{(0)} =&
	- \mathcal{C}\frac{4m_{q}^2}{\pi \omega^2}
	\cos(\pi\alpha_{0})
	,\label{eq:leading_term}\\
	Z^{(\nu)} =&
	\mathcal{C}
	\frac{m_{q}^2}{\pi} \cos(\pi\alpha_{0})
	\left(
		H_{\alpha_{-}} + H_{\alpha_{+}} -1
	\right). \label{eq:subleading_term}
\end{align}
In the standard quantization, the Dirichlet condition is imposed by $Z^{(0)}=0$.
The resulting spectrum is given by $\omega_{n} = \pm 2m_{q}\sqrt{(n+1)(n+2)}$ for $n=0,1,2,\cdots$, which is understood as a meson spectrum.\cite{Kruczenski:2003be}
Note that $Z^{(\nu)}$ remains finite in the case of $Z^{(0)} =0$ since $H_{\alpha_{-}} + H_{\alpha_{+}}$ also has a pole at $\omega = \omega_{n}$.
From eq.~(\ref{eq:Green's_function}), we obtain the Green's function as
\begin{equation}
    G^{xx}(\omega) = -\frac{\omega^2}{2}\left(
        H_{\alpha_{-}} + H_{\alpha_{+}} - 1
    \right)
    - \left(
        \frac{1}{2} + \log(m_{q} u_{\rm ct})
    \right).
\end{equation}
The residue of this function at $\omega=\omega_{n}$ is given by $Z_{n} = \mp 2 (2 n + 3)\sqrt{(n + 1)(n + 2)} m_{q}^3$ \cite{Ishigaki:2023dss}.

According to \cite{Andrade:2011dg,Andrade:2011aa}, the presence of ghosts can be checked by evaluating the Klein-Gordon inner product.
In a canonically normalized theory, this is given by
\begin{equation}
    (\phi_{\{1\}}, \phi_{\{2\}}) := i\int_{\Sigma}\dd[d]{x}\sqrt{g_{d}} n_{\mu} \left(
        \phi_{\{1\}}^{*}\nabla^{\mu}\phi_{\{2\}} - \phi_{\{2\}}\nabla^{\mu}\phi_{\{1\}}^{*}
    \right),
\end{equation}
for scalar fields, and
\begin{equation}
    (A^{\{1\}}, A^{\{2\}}) = i\int_{\Sigma}\dd[d]{x}\sqrt{g_{d}} n_{\mu} \left(
        A_{\nu}^{\{1\}*} F^{\{2\}\mu\nu}
        - A_{\nu}^{\{2\}} F^{\{1\}\mu\nu*}
    \right),
\end{equation}
for vector fields, where $\Sigma$ is a spacelike hypersurface, $g_{d}$ is the determinant of the metric on $\Sigma$, and $n^{\mu}$ is a timelike normal to $\Sigma$.
The label with braces denotes the label of the solution.
In the following, we simply take $n^{\mu} = \sqrt{-g^{tt}}\delta^{\mu}{}_{t}$.
In addition, we need to insert a dilaton-like factor $\cos^3\theta$ from the structure of the effective action (\ref{eq:Seff2}).
Considering $A = A_{y}(u) e^{-i\omega t}\dd{y}$, we obtain
\begin{equation}
\begin{aligned}
    (A^{\{1\}}, A^{\{2\}}) =& - (\omega_{1} - \omega_{2})e^{-i(\omega_{1}+\omega_{2})}
    \int\dd{u}\sqrt{-g}\cos^3\theta g^{tt}g^{yy} A_{y}^{\{1
    \}*}A_{y}^{\{2\}}\\
     = & i \frac{\omega_{1} - \omega_{2}}{\omega_{1}\omega_{2}}e^{-i(\omega_{1}+\omega_{2})}
    \int\dd{u}\frac{1-m_{q}^2u^2}{u}Z_{\{1\}}^{*} Z_{\{2\}}.
\end{aligned}
\end{equation}
Note that we omitted the subscript of the gauge invariant: $Z = Z_{2}$.
The radial integral is given by the SL inner product with coefficient (\ref{eq:SL_coeff}):
\begin{equation}
    (Z_{\{1\}}, Z_{\{2\}})_{\rm SL}:=
    \int_{0}^{1/m_{q}}\dd{u} w(u) Z_{\{1\}}^{*}(u) Z_{\{2\}}(u),
\end{equation}
which can also be evaluated by
\begin{equation}
	(Z_{\{1\}},Z_{\{2\}})_{\rm SL}=
	\left.
	\frac{p(u)}{\omega_{1}^2 - \omega_{2}^2}\left(
		Z_{\{1\}}^{*}(u)Z_{\{2\}}^{\prime}(u)
        - Z_{\{2\}}(u)Z_{\{1\}}^{\prime*}(u)
	\right)
	\right|^{1/m_{q}}_{0}.
\end{equation}
The sign of this part is related to the sign of the norm.
In general, however, this quantity includes a divergence from the AdS boundary.
In the following, we will regulate this divergence by inserting a small cut-off at $u=\epsilon$.

\begin{figure}[htbp]
\centering
\includegraphics[width=0.49\linewidth]{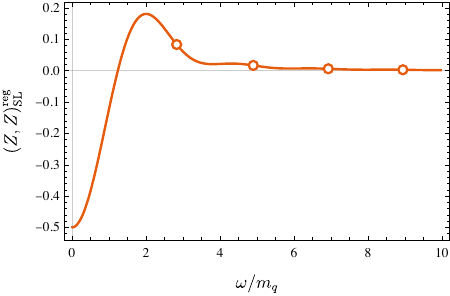}
\includegraphics[width=0.49\linewidth]{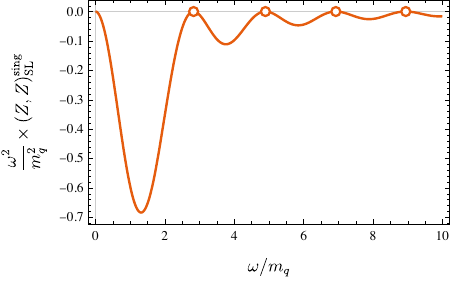}
\caption{
	Left: The regular part of the SL norm as a function of the frequency.
    Here, we set $\abs{\mathcal{C}}=1$.
   	The small circles denote the Dirichlet-modes frequencies and the values of eq.~(\ref{eq:Dirichlet_norm}).
    Right: The singular part of the SL norm multiplied by $\omega^2/m_{q}^2$ as a function of the frequency.
    The prefactor is multiplied to make the plot easier to see.
    Note that $(Z,Z)^{\rm sing}_{\rm SL}$ itself takes a non-zero value at $\omega =0$.
}
\label{fig:SL_norm}
\end{figure}

The contribution from $u=1/m_{q}$ vanishes in our case, so there is only a contribution from $u=\epsilon$, which is given by
\begin{equation}
\begin{aligned}
	\frac{(Z_{(1)},Z_{(2)})_{\rm SL}}{
        \mathcal{C}_{\{1\}}^{*}\mathcal{C}_{\{2\}}
    } =&
	\frac{-8 m_q^4}{\pi^2\omega_{1}^2 \omega_{2}^2 (\omega_{1}^2 - \omega_{2}^2)}
	\left[
		\omega _1^2 \left(
			H_{\alpha_{+}(\omega_{1})}
			+H_{\alpha_{-}(\omega_{1})}
		\right)
		-\omega _2^2 \left(
				H_{\alpha_{+}(\omega_{2})}
				+H_{\alpha_{-}(\omega_{2})}
			\right)
		\right]\\
		&\times
        \cos \left[\pi \alpha_{0}(\omega_{1})\right]
        \cos \left[\pi \alpha_{0}(\omega_{2})\right]\\
		&- \frac{16 m_q^4}{\pi ^2 \omega _1^2 \omega _2^2}
        \cos \left[\pi \alpha_{0}(\omega_{1})\right]
        \cos \left[\pi \alpha_{0}(\omega_{2})\right]
        \log(m_{q} \epsilon).
\end{aligned}
\end{equation}
We split it into two parts by $(Z_{\{1\}}, Z_{\{2\}})_{\rm SL} = (Z_{\{1\}}, Z_{\{2\}})_{\rm SL}^{\rm reg} + (Z_{\{1\}}, Z_{\{2\}})_{\rm SL}^{\rm sing} \log(m_{q} \epsilon)$.
Note that there is an ambiguity in splitting due to the choice of the logarithmic scale.
Setting $Z_{\{1\}} = Z_{\{2\}} = Z$, we obtain the SL norm as
\begin{equation}
\begin{aligned}
\frac{(Z,Z)_{\rm SL}^{\rm reg}}{\abs{\mathcal{C}}^2}
=
\frac{-2 m_q^4}{\pi ^2 \omega ^4}
\cos^2(\pi\alpha_{0})
\Bigg[&
\frac{\omega^2/m_{q}^2}{\sqrt{1+\omega^2/m_{q}^2}}
\left(
    \psi^{(1)}(\alpha_{+}+1)
    -\psi^{(1)}(\alpha_{-}+1)
\right)\\
&+4  \psi ^{(0)}\left(\alpha_{+}+1\right)
+4  \psi ^{(0)}\left(\alpha_{-}+1\right)
+8 \gamma_{\rm E}
\Bigg],
\end{aligned}
\end{equation}
and
\begin{equation}
\frac{(Z, Z)_{\rm SL}^{\rm sing}}{\abs{\mathcal{C}}^2}
=
-\frac{16 m_{q}^4}{\pi ^2 \omega ^4}
\cos ^2\left(\pi\alpha_{0}\right).
\end{equation}
Figure \ref{fig:SL_norm} shows $(Z, Z)_{\rm SL}^{\rm reg}$ and $(Z, Z)_{\rm SL}^{\rm sing}$ as functions of $\omega/m_{q}$.
Note that we do not impose any condition at $u=0$ at present but only the regularity condition at $u=1/m_{q}$.
One can see that the regular part gives a negative contribution to the norm if the mode frequency is too small.
Meanwhile, the singular part is always negative.

First, let us check the norm of the Dirichlet modes.
For $\omega=\omega_{n}$, we obtain
\begin{equation}
\begin{aligned}
\frac{(Z, Z)_{\rm SL}^{\rm reg}}{\abs{\mathcal{C}}^2} =
\frac{-\sin^2(\pi  n)}{2 \pi ^2 (n+1)^2 (n+2)^2 (2 n+3)}
\Bigg[&
(n+1)(n+2)\left(
    \psi^{(1)}(n+3)-\psi^{(1)}(-n)
\right)\\
&+(2 n+3)\left(
    \psi^{(0)}(-n)
    + \psi^{(0)}(n+3)
    + 3 \gamma_{\rm E}
\right)
\Bigg].
\end{aligned}
\end{equation}
Although $\sin(\pi n)$ vanishes for integer $n$, some terms remain finite since the digamma functions also have corresponding poles.
For the digamma functions, we can use the following Laurent expansions around negative integers $z=-n$ for $n=0,1,2,\cdots$:
\begin{equation}
	\psi^{(0)}(z) = \frac{-1}{z + n} + \cdots,\quad
	\psi^{(1)}(z) = \frac{1}{(z+n)^{2}} + \cdots.
\end{equation}
Since $n\geq0$, only the term involving $\psi^{(1)}(-n)$ remains finite.
We obtain
\begin{equation}
\begin{aligned}
\frac{(Z, Z)_{\rm SL}^{\rm reg}}{\abs{\mathcal{C}}^2} =&
\frac{1}{2 (n+1) (n+2) (2 n+3)}. \label{eq:Dirichlet_norm}
\end{aligned}
\end{equation}
The singular part $(Z,Z)_{\rm SL}^{\rm sing}$ vanishes for the Dirichlet mode frequencies.
The result confirms that the norm of the Dirichlet modes is positive definite.
From this result, the normalization constant of the Dirichlet modes can be chosen as $\mathcal{C} = \sqrt{2(n+1)(n+2)(2 n + 3)}$ \cite{Hashimoto:2014yza}.

\begin{table}
\centering
\caption{
	Neumann-mode frequencies, norm of the regular part, and singular part.
    We set $u_{\rm ct} = 1/(m_{q} e^{1/2})$ and $m_{q}=1$.
}
\label{tab:Neumann_modes}
\[
\begin{array}{c|cccc}
 n & \omega^{\prime 2}_{n} & (Z, Z)_{\rm SL}^{\rm reg} &  (Z, Z)_{\rm SL}^{\rm sing} & (Z, Z)_{\rm SL}^{\rm ren}\\
 \hline
 0 & 0. & -\frac{1}{2} & -1 & 0 \\
 1 & 17.2937 & 0.0222245 & -0.00445669 & 0.0244528 \\
 2 & 40.1456 & 0.00710861 & -0.000636429 & 0.00742682 \\
 3 & 70.9879 & 0.00314875 & -0.000168645 & 0.00323307 \\
 4 & 109.862 & 0.00166721 & -0.0000610217 & 0.00169772 \\
 5 & 156.771 & 0.000988708 & -0.000026731 & 0.00100207
\end{array}
\]
\end{table}

Next, let us consider the Neumann modes.
Their frequencies can be obtained by solving
\begin{equation}
	0 = Z^{(\nu)}
	= 
	\mathcal{C}
	\frac{m_{q}^2}{\pi} \cos(\pi\alpha_{0})
	\left(
		H_{\alpha_{-}} + H_{\alpha_{+}} -1
	\right).
\end{equation}
In the context of the double-trace deformation, this condition corresponds to considering $\lambda\to+\infty$ with a particular $u_{\rm ct}$.
The condition is reduced to $H_{\alpha_{-}} + H_{\alpha_{+}} -1 = 0$, which can be solved numerically.
In table \ref{tab:Neumann_modes}, we list the frequencies and the values of the SL norms of the first few Neumann modes.
Note that the singular part of the SL norm no longer vanishes at the Neumann-mode frequencies.
According to \cite{Andrade:2011dg}, we may need to consider a renormalized inner product.
The KG inner product of the counterterms (\ref{eq:counterterm}) manifestly cancels the logarithmic divergence in the bulk SL inner product.
The renormalized Klein-Gordon inner product should be defined as
\begin{equation}
    (A^{(1)}, A^{(2)})_{\rm ren}:=
    (A^{(1)}, A^{(2)}) + (A^{(1)}, A^{(2)})_{\rm bou},
\end{equation}
where $(A^{(1)}, A^{(2)})_{\rm bou}$ includes $\log(\epsilon/u_{\rm ct})$ as a coefficient of the kinetic term.
Correspondingly, the renormalized SL inner product is given by
\begin{equation}
\begin{aligned}
    (Z^{(1)},Z^{(2)})^{\rm ren}_{\rm SL}:=&
    (Z^{(1)},Z^{(2)})_{\rm SL} - (Z^{(1)},Z^{(2)})_{\rm SL}^{\rm sing}\log\frac{\epsilon}{u_{\rm ct}}\\
    =&
    (Z^{(1)},Z^{(2)})_{\rm SL}^{\rm reg} + (Z^{(1)},Z^{(2)})_{\rm SL}^{\rm sing}\log(m_{q} u_{\rm ct}).
\end{aligned}
\end{equation}
Thus, the value of the inner product and norm depends on the choice of $u_{\rm ct}$.
In order to avoid the negative norm of the $n=0$ mode, we have to choose $\log(m_{q} u_{\rm ct}) < -1/2$, i.e., $u_{\rm ct} < 1/(e^{1/2}m_{q})$.

\begin{figure}[htbp]
	\centering
	\includegraphics[width=0.6\linewidth]{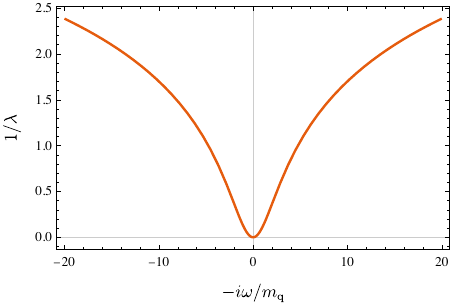}
	\caption{
        $1/\lambda$ as a function of $-i\omega/m_{q}$ in the Minkowski embedding.
		We set $u_{\rm ct} = \frac{1}{m_{q} e^{1/2}}$.
	}
	\label{fig:lambda_Minkowski}
\end{figure}

Lastly, we consider the case of the mixed boundary condition.
With the double-trace deformation, we obtain the normalized source term as
\begin{equation}
	\tilde{J}_{\rm ext} =
	\frac{1}{2} \omega^2 \left(
		\frac{2}{\lambda} - 2 \log(m_{q} u_{\rm ct}) - H_{\alpha_{-}} - H_{\alpha_{+}}
	\right).
\end{equation}
Taking $1/\lambda \to 0$ with $u_{\rm ct}=1/(m_{q} e^{1/2})$, it is reduced to the Neumann boundary condition.
Figure \ref{fig:lambda_Minkowski} shows $1/\lambda$ as a function of $-i\omega/m_{q}$ obtained by solving $\tilde{J}_{\rm ext} = 0$.
We set $u_{\rm ct} = 1/(m_{q} e^{1/2})$, i.e., $\log(m_{q}u_{\rm ct})=-1/2$.
As in the AdS$_{5}$ case, $1/\lambda$ grows logarithmically for large $-i\omega$.
This behavior implies the presence of a tachyonic mode for any positive $\lambda$.
By solving the sourceless condition for small $\omega$, we obtain the lowest approximation as $\omega \approx -8 m_{q}^2/\lambda$.
On the other hand, since $1/\lambda$ is bounded below, it seems to be possible to avoid a tachyonic mode by considering negative $\lambda$ in this case.

From the boundary action (\ref{eq:Sfin}), the renormalized SL inner product should be modified as
\begin{equation}
    (Z_{\{1\}},Z_{\{2\}})^{\rm ren}_{\rm SL}:=
    (Z_{\{1\}},Z_{\{2\}})_{\rm SL}^{\rm reg} + (Z_{\{1\}},Z_{\{2\}})_{\rm SL}^{\rm sing}\left(
        \log(m_{q} u_{\rm ct}) - \frac{1}{\lambda}
    \right).
\end{equation}
Meanwhile, we can consider the response function defined by (\ref{eq:response_function_chi}) with finite $\lambda$.
In the present case, the response function takes the form
\begin{equation}
    \chi^{xx}(\omega) = \sum_{n=0}^{\infty} \frac{\mathcal{Z}_{2,n}}{\omega^2 - \omega^2_{\lambda,n}}
    + \text{(regular part)},
\end{equation}
where $\omega_{\lambda,n}$ is a mode frequency, and $\mathcal{Z}_{2,n}$ is the residue at $\omega^2 = \omega_{\lambda,n}^2$.
In our convention, physical modes should have negative $\mathcal{Z}_{2,n}$ whereas ghosts may have a positive residue.
We show the mode frequencies, residues, and values of the renormalized SL norm for $\lambda = 2$ and $-2$ in table \ref{tab:mixed_modes}.
As it was expected from figure \ref{fig:lambda_Minkowski}, a tachyon exists for $\lambda =2$ that is also a ghost.
The signs of the residues correspond to the signs of norm correctly.
Since the pole of the tachyon is located on the imaginary axis, it may not affect the positivity of the spectral function even though this is a ghost.
For $\lambda = -2$, the tachyon disappears, and there is no ghost.

\begin{table}[htbp]
\centering
\caption{
    Modes frequencies, residues, and values of the SL norm under the mixed boundary condition for $\lambda = 2$ and $-2$.
}
\label{tab:mixed_modes}
\begin{subtable}{0.45\linewidth}
\caption{$\lambda = 2$}
\[
\begin{array}{c|ccc}
 n&\omega_{\lambda, n}^2 & \mathcal{Z}_{2,n} & (Z,Z)_{\rm SL}^{\rm ren}\\
 \hline
 0 & -6.34082 & 30.5445 & -4.72587 \\
 1 & 15.5481 & -14.459 & 0.027712 \\
 2 & 38.1435 & -43.5508 & 0.0078756 \\
 3 & 68.809 & -87.0523 & 0.00335075 \\
 4 & 107.535 & -146.12 & 0.00174017
\end{array}
\]
\end{subtable}
\begin{subtable}{0.45\linewidth}
\centering
\caption{$\lambda=-2$}
\[
\begin{array}{c|ccc}
 n & \omega_{\lambda, n}^2 & \mathcal{Z}_{2,n} & (Z,Z)_{\rm SL}^{\rm ren}\\
 \hline
 0 & 2.80269 & -2.70118 & 0.149118 \\
 1 & 18.6596 & -10.8333 & 0.0225613 \\
 2 & 41.6161 & -25.899 & 0.00713446 \\
 3 & 72.5713 & -48.8105 & 0.0031524 \\
 4 & 111.556 & -80.622 & 0.00166774
\end{array}
\]
\end{subtable}
\end{table}

\section{Conclusion and Discussions}\label{sec:discussions}
In this paper, we closely examine the influence of the scheme-dependent regularization required in AdS$_5$ spacetime on the double-trace deformation for the gauge fields.
Our analysis using both analytic and numerical approaches suggests that a naive choice of a dimensionful scale, which is chosen arbitrarily in the regularization procedure, unavoidably induces instabilities in the system, confirmed by computing the (quasi-)normal modes.
This instability originates from the logarithmic dependence of the fields in asymptotically AdS$_{d+1}$ spacetimes for even $d$, and has received little attention in previous studies because of the choice of parameters and observables.
This instability does not appear for odd $d$, such as AdS$_{4}$ spacetime as shown in appendix \ref{appendix:SAdS4}.

In the five-dimensional Einstein-Maxwell theory with and without a charge density, which have an asymptotically AdS$_{5}$ spacetime as a solution, we find that an unstable mode inevitably appears for any value of the coupling constant $\lambda$. In other words, $1/\lambda$ vanishes in the region $-i\omega > 0$, i.e., $\lambda$ diverges at a particular positive imaginary frequency as shown in figure \ref{fig:lambda_SAdS5} and figure \ref{fig:lambda_RNAdS5}.
The purely imaginary frequency $\omega_{\rm I} = -i\omega$ can be roughly identified with the spatial momentum, so this can be rephrased as the statement that $\lambda$ diverges at a particular momentum.

Due to the running of the coupling, the coupling constant generally diverges at a specific RG-invariant scale, which is known as the Landau pole and corresponds to a breakdown of perturbation theory.
The instability we study seems to be related to this RG-invariant scale as we demonstrate in \eqref{eq:predict} and as discussed in \cite{Faulkner:2012gt}.
{
From the perspective of the quantum field theory with $U(1)$ gauge symmetry, this breakdown is a natural consequence.
As is well known in QED, such a theory is generally UV incomplete.
Since the double-trace deformation is adding the kinetic term of the $U(1)$ gauge fields on the boundary, the dual theory is analogous to QED, and the emergence of tachyonic and ghost mode can be understood as evidence of the corresponding UV incompleteness.
Our results therefore indicate that the method of the double-trace deformation breaks down at high energies due to the above reasons, when applying to AdS$_5$ cases.
}

As mentioned in \cite{Mauri:2018pzq} and reviewed around (\ref{eq:RPA}) in this paper, the modes in the presence of the double-trace deformation correspond to the poles of the RPA-like response function. The RPA formula includes only direct-chain diagrams (figure \ref{fig:RPA_diagram}), so this may also be related to the origin of the problem. It remains an open question whether additional contributions beyond this approximation could modify the structure of the collective modes.

The mode analysis of the Minkowski embedding in the D3-D7 model, performed in section \ref{sec:D3-D7_Minkowski}, shows that the tachyon appears for positive $\lambda$, in the same manner as in the AdS$_{5}$ case.
From the analysis of the SL norm and the residue of the response function, this mode is not only a tachyon but also a ghost.
The tachyons in other cases may also be thought of as similar ghost-like modes.
On the other hand, we find that there is no tachyon or ghost for negative $\lambda$ in the case of the Minkowski embedding.
Although we focus on the case of zero temperature, we can introduce finite temperature into the D3-D7 model by considering SAdS$_5\times S^5$ background.
If we increase the temperature sufficiently, the D7-brane falls into the black hole horizon, which is called black hole embedding.
In the black hole embedding, we expect to obtain results similar to those in the SAdS$_5$ case, and the tachyon will appear even for negative $\lambda$, regardless of the value of $u_{\rm ct}$.

In this paper, we have focused mostly on the bulk Maxwell theory, but it is also natural to consider more complex models.
For example, one may consider a model with higher-derivative terms.
Higher-derivative terms contribute as higher powers of momentum in the mode equation and affect the dispersion relations of the modes.
One may also consider a model with non-linear terms of the gauge fields.
In fact, the DBI action of the D3-D7 model studied in Section \ref{sec:D3-D7_Minkowski} is one example of a non-linear model.
However, in our case, since only linear perturbations are relevant to the mode analysis, the non-linear terms do not contribute.
The non-linear terms become important when the gauge fields have background profiles, such as in the case with finite charge density.
The instability problem we observed may be resolved in these models with higher-derivative and non-linear terms.

The authors of \cite{Andrade:2011aa} argued that tachyons and ghosts can be removed by considering appropriate boundary terms, i.e., UV completions, in the case of scalar fields with non-normalizable modes.
This is achieved by introducing a finite mass term in the boundary.
In our case of the vector fields, however, including such a mass term breaks the gauge symmetry.
One possible modification that preserves the gauge symmetry is to consider a higher-derivative term, such as $\nabla^2 F^2$, in the boundary action.
Clarifying such possible UV completions and their implications for dynamical gauge fields would be an interesting direction for future work.


\section*{Acknowledgments}
The work of SI is supported by NSFC, China (Grant No. W2433015).
The work of MM is supported by Shanghai Sci-tech Co-research Program (Grant No. 25HB2700600).
The authors thank Shin Nakamura, Chanyong Park, and Hyun-Sik Jeong for useful discussion.

\appendix

\section{Examples in the Schwarzschild \texorpdfstring{AdS$_{4}$}{
AdS4} spacetime}\label{appendix:SAdS4}
In this appendix, we discuss the double-trace deformation for the gauge fields in the Schwarzschild AdS$_{4}$ spacetime.
The metric is given by
\begin{equation}
	\dd{s}^2 = \frac{-f(u) \dd{t}^2 + \dd{\vec{x}}^2}{u^2} + \frac{\dd{u}^2}{f(u) u^2},\quad
	f(u) = 1 - \frac{u^3}{u_{0}^3}.
\end{equation}
where $\vec{x}=(x,y)$, and the dual field theory is in the (2+1)-dimensional spacetime. Here, we consider the probe limit for simplicity.
Then, the action is given by the Maxwell theory in the Schwarzschild AdS$_{4}$ spacetime:
\begin{equation}
	S = \int \dd[5]{x} \sqrt{-g}\left(
		- \frac{1}{4} F_{\mu\nu} F^{\mu\nu}
	\right).
\end{equation}
As in the main text, we consider small perturbations
\begin{equation}
	A = a_{t}(t,x,u) \dd{t} + a_{x}(t,x,u) \dd{x} + a_{y}(t,x,u) \dd{y}. 
\end{equation}
The gauge invariants are defined as
\begin{equation}
	Z_{1} = i\omega a_{x}(u) + i k_{x} a_{t}(u),\quad
	Z_{2} = i\omega a_{y}(u),
\end{equation}
in Fourier space, corresponding to the longitudinal and transverse sectors, respectively.
The equations of motion are
\begin{align}
	0=&
	Z_{1}''(u) + 
		 \frac{\omega^2 f'(u)}{(\omega^2 - k_{x}^2 f(u))f(u)}
	    Z_{1}'(u) + \frac{\omega^2 - k_{x}^2 f(u)}{f(u)^2} Z_{1}(u),\\
	0 =&
	Z_{2}''(u) +
		 \frac{f'(u)}{f(u)}
	Z_{2}'(u) + \frac{\omega^2 - k_{x}^2 f(u)}{f(u)^2} Z_{2}(u).
\end{align}
Note that the difference with the case of AdS$_5$ is the absence of $Z'/u$ term.
From the equations of motion, these fields are asymptotically expanded near the AdS boundary as
\begin{equation}
	Z_{\alpha} = Z_{\alpha}^{(L)} 
	+ Z_{\alpha}^{(S)} u + \order{u^2},
\end{equation}
where $Z_{\alpha}^{(L)}$ and $Z_{\alpha}^{(S)}$ are the leading and sub-leading coefficients, respectively. Note that there is no logarithmic term in the asymptotic behavior near the AdS boundary in the case of AdS$_{4}$.

For $k_{x}=0$, the equations for the longitudinal and transverse fields are equal, and the solution can be written analytically as
\begin{align}
    Z(u) =&c_{1} \cos \left[ \frac{u_{0}\omega}{6}\left( 2\sqrt{3} \arctan \frac{2u + u_{0}}{\sqrt{3}u_{0}} + \log \frac{u^{2} + u_{0}u + u_{0}^{2}}{(u-u_{0})^{2}}\right) \right] \\ &+ c_{2} \sin \left[ \frac{u_{0}\omega}{6}\left( 2\sqrt{3} \arctan \frac{2u + u_{0}}{\sqrt{3}u_{0}} + \log \frac{u^{2} + u_{0}u + u_{0}^{2}}{(u-u_{0})^{2}}\right) \right],
\end{align}
where $c_{1}$ and $c_{2}$ are constants. In the vicinity of the black hole horizon $u\sim u_{0}$, the solution behaves as
\begin{eqnarray}
    Z(u\sim u_{0}) &=& c_{1}\cos\left( \frac{u_{0}\omega}{3} \log (u_{0}-u) \right) -c_{2}\sin\left( \frac{u_{0}\omega}{3} \log (u_{0}-u) \right), \nonumber \\
    &=& (c_{1} + i c_{2})(u_{0}-u)^{i \frac{u_{0}}{3}\omega} + (c_{1} - i c_{2})(u_{0}-u)^{-i \frac{u_{0}}{3}\omega},
\end{eqnarray}
where the first and second terms correspond to the outgoing and ingoing solutions, respectively. For our purpose, we choose the ingoing solution, i.e. $c_{1}=-i c_{2}$.
Thus, we obtain the ingoing solution as
\begin{equation}
    Z(u) = C \exp \left[ \frac{iu_{0}\omega}{\sqrt{3}}
    \left(
    {\rm arccot} \frac{\sqrt{3}u_{0}}{2u + u_{0}} + \log \frac{u^{2} + u_{0} u+ u_{0}^{2}}{(u-u_{0})^{2}}
    \right)
    \right],
\end{equation}
where $C$ is a constant.
Near the AdS boundary, the solution is expanded as
\begin{equation}
    Z(u) = Ce^{i\frac{\pi u_{0}\omega}{6\sqrt{3}}} \left( 1 + i \omega u + \cdots \right).
\end{equation}

In the case of AdS$_{4}$, the boundary condition with the double-trace deformation is given by
\begin{equation}
    J_{\rm ext} = - \frac{\omega}{\lambda} Z^{(L)} - \frac{\omega}{\omega^{2} - k_{x}^{2}}Z^{(S)}=0.
\end{equation}
Considering the case of $k_{x}=0$, the $1/\lambda$ as a function of $\omega$ is written as
\begin{equation}
    \frac{1}{\lambda} = -\frac{1}{\omega^{2}}\frac{Z^{(S)}}{Z^{(L)}} = \frac{1}{i \omega} = -\frac{1}{\omega_{\rm I}}. \label{eq:lambda_inverse_AdS4}
\end{equation}
As discussed in \cite{Ahn:2022azl}, this implies that the overdamped mode $\omega = - i \lambda$ emerges in the case of the dynamical gauge field, whose relaxation time corresponds to $\tau = 1/\lambda$.
As a result, the emergence of unstable modes implies a negative value of $\lambda$, corresponding to a negative permittivity that leads to a trivial instability, while no instability appears for any positive value of $\lambda$.
Since no logarithmic counterterm is required in the case of AdS$_4$, the result is free from any scheme dependence \cite{Mauri:2018pzq}.

The Green's function is simply given by
\begin{equation}
    G^{xx}(\omega) = \frac{Z^{(S)}}{Z^{(L)}} = i \omega,
\end{equation}
and the resulting response function from \eqref{eq:response_function_chi} becomes
\begin{equation}
    \chi^{xx}(\omega) = \frac{i \omega}{ 1 + \frac{i \lambda}{\omega}}.
\end{equation}
As expected, the pole of the response function corresponds to the overdamped mode we obtained above: $\omega = - i \lambda$.

\bibliography{main}
\bibliographystyle{JHEP}
\end{document}